\documentclass[amssymb,aps,twocolumn,showpacs, superscriptaddress,nofootinbib]{revtex4-1}
\usepackage[]{graphicx}
\usepackage[]{amsmath}
\usepackage{hyperref}
\usepackage{color}

\def\ket#1{| #1 \rangle}

\def\kb#1#2{| #1 \rangle\!\langle #2 |}

\def\cV{\mathcal{V}}

\DeclareMathOperator*{\argmin}{arg\,min}
\def\eq#1{Eq.~\eqref{eq:#1}}
\def\fig#1{Fig.~\ref{fig:#1}}
\def\sec#1{Sec.~\ref{sec:#1}}

\begin{document}

\title{Self-correction in Wegner's 3D Ising lattice gauge theory}
\author{David Poulin}
\email{David.Poulin@USherbrooke.ca}
\affiliation{D\'epartement de Physique \& Institut Quantique, Universit\'e de Sherbrooke, Qu\'ebec, Canada}
\affiliation{Canadian Institute for Advanced Research, Toronto, Ontario, Canada M5G 1Z8}
\author{Roger G. Melko}
\email{rgmelko@uwaterloo.ca}
\affiliation{Department of Physics and Astronomy, University of Waterloo, Ontario, N2L 3G1, Canada}
\affiliation{Perimeter Institute for Theoretical Physics, Waterloo, Ontario N2L 2Y5, Canada}
\author{Matthew B. Hastings}
\email{mahastin@microsoft.com}
\affiliation{Station Q, Microsoft Research, Santa Barbara, CA 93106-6105, USA}
\affiliation{Quantum Architecture and Computation Group, Microsoft Research, Redmond, WA 98052, USA}

\date{\today}

\begin{abstract}
Motivated by the growing interest in self-correcting quantum memories, we study the feasibility of self-correction in classical lattice systems composed of bounded degrees of freedom with local interactions. We argue that self-correction, including a requirement of stability against external perturbation, cannot be realized in system with broken global symmetries such as the 2d Ising model, but that systems with local, i.e. gauge, symmetries have the required properties.  Previous work gave a three-dimensional {\it quantum} system which realized a self-correcting {\it classical} memory \cite{DKLP02a}.  Here we show that a purely classical three dimensional system, Wegner's 3D Ising lattice gauge model \cite{W71a}, can also realize this self-correction despite having an extensive ground state degeneracy.  We give a detailed numerical study to support the existence of a self-correcting phase in this system, even when the gauge symmetry is explicitly broken. More generally, our results obtained by studying the memory lifetime of the system are in quantitative agreement with the phase diagram obtained from conventional analysis of the system's specific heat \cite{PhysRevB.82.085114}, except that self-correction extends beyond the topological phase, past the lower critical temperature.
\end{abstract}

\maketitle

\section{Introduction and motivation}

A self-correcting memory is a passive physical device that stores information robustly at finite temperature despite fluctuations of its external parameters like magnetic field, pressure, etc. Such a system can be prepared in one of a finite number of initial states, and the identity of this initial state can in principle be reconstructed with high probability up to some mixing time, after which all signatures of the initial configuration is lost.  We say that the system is self-correcting if the mixing time grows with the system size. In particular, we are interested in self-correcting memories that arise from systems composed of localized, bounded degrees of freedom with short-range interactions. 

It is  well established that reliable computation can be realized from unreliable components using fault-tolerant techniques both in the classical \cite{vN56a,DO77a,P85a,S96b} and quantum \cite{Sho96a,Ste96c,AB97a,Pre98a,KLZ98a,Kit03a,Kni05a} settings. Moreover, fault-tolerant computation can be realized from local cellular automaton \cite{T80a,BCG82a,G88a}, which in physical terms corresponds to a lattice system with local interactions that are periodic in time. However, these processes dissipate heat. In contrast, a self-correcting memory is stabilized thermodynamically---i.e., through its interaction with a heat reservoir---and does not require external power. 

To serve as a memory, a system must possess more than one metastable state, i.e., it must display phase coexistence. Information is stored in the system by preparing one of these states, and the probability of information corruption, i.e., the probability that the system spontaneously transitions from one phase to another, is exponentially suppressed with the system size. 

Robust phase co-existence is however ruled out by the Gibbs phase rule \cite{G61a}: for a system with $N$ external parameters, the coexistence of $P$ stable phases can only occur in a sub-manifold of codimension $P-1$. In other words, phase co-existence requires fine-tuning of the system's parameters. As we will explain below, this rule can easily be understood in the Landau-Ginzburg paradigm of local order parameters and spontaneously broken symmetries, where it also rules out the existence of self-correcting memories. In this article, we show how the Gibbs phase rule can be circumvented by turning to systems with non-local order parameters.

The motivations to study this problem stem from many sources. In the classical setting, the question of emerging global order from local noisy interactions has a long history. For instance, the phase rule forbids the existence of a phase transition in a one-dimensional, translationally-invariant system composed of bounded, discrete degrees of freedom. Yet, G\'acs' cellular automaton \cite{Gac86a,G01b} provides a ``counter-example'' to this rule, that is made possible using a time-dependent periodic Hamiltonian instead of a constant Hamiltonian \cite{BG85a}. Here we present an alternative way of escaping the phase rule.

In the quantum setting, the prospect of quantum technologies has generated a growing interest for robust quantum memories. While the theory of fault-tolerant quantum computation is well developed, a self-correcting quantum memory \cite{Bac05a} would enable passive, reliable quantum information storage, analogously to classical hard drives. The quantum model of \cite{Kit03a,DKLP02a} realizes a self correcting quantum memory in four spacial dimensions and a classical memory in three spatial dimensions, but leaves open the questions of a robust classical memory in a classical system and of a robust quantum memory in lower spatial dimensions. 

Lower-dimensional quantum model systems have been proposed \cite{Bac05a,HCC08a,HWRL12a,PHWL12a,M14a,B16b} and disputed \cite{HP10a,LP13a,LYPP15a}, so the existence of a self-correcting quantum memory in less than four spatial dimensions remains an open question to date. Given this status, it seams reasonable to step back and study classical self-correcting memories in a classical system before turning to the quantum setting. In particular, the study of self-correcting quantum memories in two spatial dimensions \cite{Bac05a,HCC08a,HWRL12a,PHWL12a} appears premature given that our construction of a classical self-correcting memory is the only model we are aware of and requires three spatial dimensions. 

In addition, the conditions for a quantum memory are more stringent than those for a classical memory. Suppose the energies of two memory states 0 and 1 differ by some unknown, possibly fluctuating quantity $\Delta$. If $\Delta$ is extensive -- i.e., if there is a constant energy density difference between the two states -- then thermal fluctuations will drive the memory to the state of lower energy. This thermal instability is a problem for both quantum and classical memories. However, a quantum memory must not only preserve the discrete memory states $0$ and $1$, it must also preserve coherent superpositions thereof $\ket{\psi(0)} = \alpha\ket 0 + \beta \ket 1$.  For any finite energy splitting $\Delta$ between the two states, Schr\"odinger's time evolution will introduce an unknown phase over time $\ket{\psi(t)} = \alpha\ket 0 + e^{-i\Delta t/\hbar}\beta \ket 1$,  effectively destroying the quantum superposition. Such a dephasing problem only affects quantum memories and does not require an extensive energy difference $\Delta$, showing that the criteria to realize a self-correcting quantum memory are much more stringent than for a classical memory. In contrast, quantum mechanical models are richer than classical models in some fixed spatial dimension, so the question of classical self-correction in classical systems is a non-trivial intermediate case to study.  

The rest of this article is organized as follows. In the next section, we argue that systems with a broken local symmetry cannot realize a self-correcting memory. So in \sec{gauge}, we turn to systems with a gauge symmetry and explain how the previous argument breaks down. Section \ref{sec:model} presents a detailed model and  explain why it is stable against external perturbations, thermal fluctuations, and both. Numerical simulations of that model are presented in \sec{numerics} where the self-correcting behavior is clearly observed. We conclude with general remarks and some open questions.

\section{Symmetry broken phase}

To build intuition, consider a two-dimensional ferromagnetic  Ising model with Hamiltonian $H = -J\sum_{\langle i,j\rangle} \sigma_i\sigma_j$, where spins take values $\sigma_j\in \{-1,+1\}$, (see e.g. \cite{PB94a}). It has two ground states of fully polarized all-up and all-down spins and this degeneracy can be used to store one bit of information. At non-zero temperature, thermal fluctuations will create droplets of inverted spins, reducing magnetization. However, the energy of an error droplet $D$ is proportional to the length of its boundary $E= 2J|\partial D|$, so at temperature $T$, large droplets are suppressed by a Boltzmann factor $e^{-E/k_BT}$ (we set $k_B = 1$). Below the critical Curie temperature $T_C$,  error droplets are  typically too sparse to percolate, so the sign of the magnetization retains information about the stored bit of information. Statistical fluctuations could lead to spontaneous percolation of the droplets, but such fluctuations are exponentially suppressed with the system size, meaning that the memory lifetime grows exponentially with the system size.

This stability is attributable to a symmetry -- flipping all the spins leaves the Hamiltonian invariant. Sub-critical temperature thermal states spontaneously break this symmetry, either choosing a positive or negative magnetization sector, thus enabling the system to serve as a memory. This requires fine-tuning however: generic perturbations to the Hamiltonian will break this symmetry and favor one sector over the other. For instance, a magnetic field will add a term $B\sum_i \sigma_i$ to the Hamiltonian. The energy of a spin-down droplet $D$ in a spin-up background is modified to $E = -2B|D|+2J|\partial D|$. Because the area $|D|$ of a droplet grows faster than its boundary $|\partial D|$, no matter how small the magnetic field $B$ is, the Boltzmann factor will favor the formation of large droplets, and hence  a unique stable sector. We see that phase co-existence is restricted to a codimension 1 manifold of the $(B,T)$ phase diagram in accordance to Gibbs' phase rule. 

Because of the required fine-tuning, the Ising model is not a robust phase of matter -- it is merely a symmetry-protected phase of matter. A self-correcting memory is a symmetry broken finite-temperature phase, robust to generic physical perturbations that are not constraint by any symmetries. This is by definition impossible to realize in  symmetry-broken phases with a {\em local} order parameter. Suppose indeed that Hamiltonian $H$ has a spontaneously broken symmetry with associated order parameter $M$, i.e.,  $M$ takes distinct expectation values in the different thermal sectors. Adding a perturbation $\mu M$ to the Hamiltonian will favor one sector over the others, so the system will thermalize to this unique sector irrespective of initial conditions. Moreover, because $M$ is a local order parameter, the term $\mu M$ consists of physically realistic local interactions. In the Ising model for instance, $M$ would be magnetization and the perturbation would correspond to an external magnetic field.

\section{Non-local order}
\label{sec:gauge}
%
We have just argued that in the presence of symmetry-breaking perturbations, thermal stability is incompatible with the existence of a local order parameter. So we turn to systems with non-local order parameters. In the quantum setting, it is possible for a system to possess two distinct ground states $\ket{\psi_0}$ and $\ket{\psi_1}$ that are locally indistinguishable. The two state 
\begin{align}
\ket{\psi_0^{AB}} &= \frac 1{\sqrt 2}(\ket{\!\uparrow^A\uparrow^B} + \ket{\!\downarrow^A\downarrow^B}) \quad {\rm and}\\
\ket{\psi_1^{AB}} &= \frac 1{\sqrt 2}(\ket{\!\uparrow^A\uparrow^B} - \ket{\!\downarrow^A\downarrow^B})
\end{align}
illustrate this idea: in both states, the  single-spin state obtained from a partial trace is maximally mixed 
\begin{equation}
\rho^A_{0/1} = \rho^B_{0/1} = \frac 12 (\kb{\!\uparrow}{\uparrow\!} + \kb {\!\downarrow}{\downarrow\!}),
\end{equation}
 yet the two states are orthogonal to each other. Thus, any single-spin measurement cannot distinguish between the two states. Topologically ordered systems extend this idea to local Hamiltonians whose degenerate ground states are indistinguishable from each other given any observable acting on a homologically trivial region of the system manifold space. Thus, these systems naturally present a stability to local perturbations \cite{BHM10a,K10c,BH11b}.

In the classical setting, two distinct configurations of a system composed of local degrees of freedom must unavoidably differ locally, so are subject to an energy splitting by a local field. To obtain a self-correcting memory, we therefore cannot encode information states in distinct system configurations. Instead,  we choose a one-to-many encoding where each information state is encoded in an ensemble of classical states, e.g., 0 is encoded in any configuration from the ensemble $\Omega_0 = \{\sigma_0^1,\sigma_0^2,\ldots\}$ and 1 is encoded in any configuration from the ensemble $\Omega_1 = \{\sigma_1^1,\sigma_0^1,\ldots\}$. While any of these individual states can be locally distinguished, the distinct ensembles can be chosen to be statistically locally indistinguishable. 

A concrete way of realizing this encoding uses a Hamiltonian with a gauge symmetry. The spectrum of such a model is exponentially degenerate because configurations related by a gauge transformation have the same energy. The information ensembles $\Omega_j$ will consist of sets of states that are related by a gauge transformation. The fact that these ensembles cannot be distinguished locally is then a corollary of Elitzur's theorem \cite{Eli75a,Kog79a} which states that a gauge symmetry cannot be spontaneously broken. Moreover, the non-local order parameter characterizing the gauge model remains well defined in the absence of a gauge symmetry, so it can continue to serve as the information read-out in the model even when a symmetry-breaking field in added. The next section illustrates these ideas with a concrete model, Wegner's 3D Ising lattice gauge theory \cite{W71a}.

\section{Model}
\label{sec:model}

\begin{figure}
\includegraphics[width=0.3\textwidth]{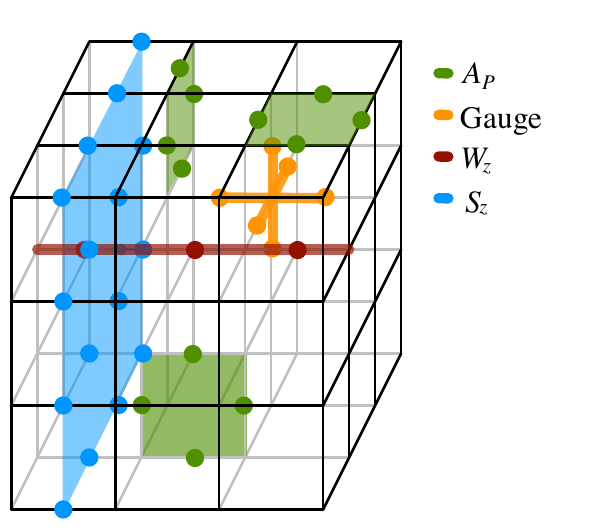}
\caption{The model consist of Ising spins $\sigma_i$ placed on each {\em edge} of a regular square lattice with periodic boundary conditions. Green plaquettes represent the coupling terms of the Hamiltonian. Orange vertices represent gauge symmetries. The red string is a gauge invariant Wilson loop $W_z$. Flipping all spins along the blue plane $S_z$ perpendicular to the $z$ axis inverts the value of $W_z$.}
\label{fig:Lattice}
\end{figure}

The model we consider is a cubic lattice of size $L\times L\times L$, with Ising spins $\sigma_i \in \{-1,+1\}$ residing on each {\em edge}, see \fig{Lattice} for an illustration. The Hamiltonian is the sum over plaquettes 
\begin{equation}
H_0 = -J \sum_P A_P
\label{eq:H}
\end{equation} 
where $J>0$ and  for plaquette $P$ with boundary edges $\partial P = (i,j,k,l)$, we define $A_P = \sigma_i\sigma_j\sigma_k\sigma_l$. For simplicity, we assume the lattice has periodic boundary conditions. Beyond its use in lattice gauge theory \cite{Kog79a}, this model is familiar in quantum error-correction as it consists of the``classical part"  of the  three-dimensional toric code \cite{Kit03a,DKLP02a}. That quantum model is known to be a self-correcting classical memory; the question we ask here is whether it retains that property without the kinetic quantum terms. Note that the spectrum of the model changes drastically with the presence of the quantum kinetic term. In particular, the quantum model has some constant ground-state degeneracy while the degeneracy of the classical model is extensive. Thus, there is a priori no reason to assume any relation between the thermal properties of the two models, particularly in the presence of an external field as we explain later.

This Hamiltonian has a gauge symmetry generated by flipping all six spins adjacent to a vertex. To see that this is a symmetry, note that a plaquette and a vertex share an even number $\eta$ (either 0 or 2) of edges, so the gauge transformation will change a plaquette coupling $A_P$ by a multiplicative  factor $(-1)^\eta = 1$. In other words, this follows from the fact that the boundary of a plaquette has itself no boundary, $\partial\partial P = 0$. 

\subsection{observables}

A gauge-invariant observable is obtained by considering the product of the spins along any closed loop $\ell$, generating a so-called Wilson loop $W_\ell = \prod_{i\in \ell} \sigma_i$. The gauge invariance follows from $\partial\partial \ell = 0$. For loops $\ell$ that are boundary of surfaces $S$, i.e. $\ell = \partial S$,  the corresponding Wilson loops are obtained by taking the product of the enclosed plaquette terms $W_{\partial S} = \prod_{P\in S}A_P$, so they are coupled to energy density in $S$. High-temperature expansions can be used to show \cite{W71a,Kog79a} that $\langle W_\ell \rangle \sim e^{-\alpha |S|}$ where $S$ is the minimal surface area enclosed by the loop, i.e., for which $\partial S = \ell$. At low temperature however, any single spin-flip along $\ell$ will  invert the sign of $W_\ell$, so the expectation of a Wilson loop vanishes exponentially with its length  $\langle W_\ell \rangle \sim e^{-\gamma |\ell|}$ where to leading order $e^{-\gamma}$ is the probability of an unflipped spin, so $e^{-\gamma} \approx (1-e^{-4J/T})$.  This discrepancy between high- and low-temperature calculations suggest the existence of a finite-temperature phase transition. Indeed, this model is dual to a 3D Ising model \cite{W71a,Kog79a} whose critical temperature is well established numerically $T_c \approx 1.314$ \cite{GT96a}.

A system embedded on a topologically non-trivial  manifold---such as a cube with periodic boundary conditions or a punctured cube---admits homologically non-trivial loops $\ell \neq \partial S$ to which the above high-temperature expansion does not apply. These Wilson loops are gauge-invariant observables that are decoupled from energy. One such example, illustrated on \fig{Lattice} is the loop which winds around the three-dimensional torus in the $z$ direction $W_z$.

To see that $W_z$ is indeed decoupled from the energy, consider the dual plane $S_{z}$ which consists of all the edges in the $z$ direction with a fixed $z$ coordinate, see \fig{Lattice}. Inverting all the spins in $S_{z}$ leaves all plaquette terms $A_P$ invariant because $S_{z}$ intersects $A_P$ on an even number of edges. On the other hand, $W_z$ and $S_{z}$ intersect on an odd number edges, so flipping all spins in $S_{z}$ inverts the value of $W_z$. We conclude that, given any spin configuration $\sigma$, there exist another  configuration $\sigma'$ obtained by flipping all the spins in $S_{z}$ such that $H_0(\sigma) = H_0(\sigma')$ and $W_z(\sigma) = -W_z(\sigma')$. We encode one bit of information using the two ensembles of spin configurations $\Omega_+$ and $\Omega_-$ consisting of the ground states of $H_0$ with $W_z = + 1$ and $W_z = -1$ respectively. 

\subsection{Stability against perturbation}

To understand the robustness of this phase, let us repeat the above argument in the presence of a generic perturbation $V = \sum_i v_i$ to the Hamiltonian $H = H_0+V$, where each term $v_i$ is a bounded function $|v_j|\leq K$ of all the spins within a constant radius $r$ away from spin $i$.  Given a spin configuration $\sigma$ of energy $H(\sigma)$, consider the configuration $\sigma'$ obtained by flipping the spins in $S_{z}$. It is easy to see that the two configurations have a vanishing energy density difference as the volume $\cV = L^3$ grows
\begin{equation}
\frac{|H(\sigma)-H(\sigma')|} \cV = \frac{|V(\sigma)-V(\sigma')|} \cV \leq \frac{4Kr}{\cV^{1/3}}.
\end{equation}
In the thermodynamic limit, thermal fluctuations will thus not discriminate between the two sectors, so there is phase co-existence. 

\subsection{Thermal stability}

Thermal stability is more subtle. Suppose we prepare a ground state $\sigma$ of $H_0$ with a fixed value of $W_z = w$ and let the system thermalize. In the ground state, we have $A_P = 1$ for all $P$, but in thermal equilibrium some of the $A_P$ will take value -1. Excited states are obtained from the ground state by flipping spins contained in dual-membranes -- two-dimensional sub-manifolds akin of the region $S_{z}$ but which don't span an entire plane. Just like in the two-dimensional Ising model, the energy of such an error membrane $M$ grows proportionally to its boundary $E = 2J|\partial M|$, so below a critical temperature the membranes are confined. Despite this confinement, however, at any non-zero temperature, we expect a constant density of membranes. Since the sign of $W_z$ is inverted every time it intersects an error membrane,  $W_z$ averages to 0 in the thermodynamic limit as $1-e^{-\gamma L_z}$ where $e^{-\gamma}$ is roughly the density of error membranes. 

Despite the vanishing of $W_z$, the system retains some information about its initial configuration. After letting the system interact with a heat bath for some time, suppose we were to cool it down. Error membranes would slowly shrink and disappear, returning the system to a spin configuration with the same value $W_z=w$ as initially, unless the thermalizing and cooling processes have generated a membrane that spans a homologically non-trivial  plane. Because error membranes are confined, the probability of such an event vanishes in the thermodynamic limit. 

While cooling the system back to $T=0$ is not physically possible, it shows that in principle, the information is still present in the system. In fact, the cooling phase does not need to be implemented to read out the information: it is used here only to illustrate that in principle the information is still encoded in the system, although in a hidden form. There exists an alternative ``algorithmic'' procedure for retrieving the information. It consists of reading out the thermal spin configuration $\sigma$ entirely, recording the value of each individual spin. From this knowledge, the value of each $A_P$ can be computed. The $P$ with $A_P=-1$ indicate the boundary $\ell$ of error membranes. Then, an algorithmic procedure called {\em decoding} can be used to determined what are the smallest membranes consistent with these boundaries: $\argmin_M \{|M|:\partial M = \ell\}$. This may be computationally expensive but can certainly be done in principle. Inverting the spins contained in these minimal membranes should return the system to its initial value of $W_z$ unless some error membranes had grown and percolated to reach a macroscopic size.  Thus, the self-correcting phase truly  corresponds to a membrane confinement phase, so we expect the self-correcting phase diagram to coincide with the one obtained for confinement, which is dual to the 3D Ising model $h=0$.

\subsection{General stability}
\label{sec:GS}

Also, we can consider the thermal stability in the presence of a gauge symmetry-breaking field. Suppose a  magnetic  field $V = -B\sum_j\sigma_j$  is added, energetically favoring $\sigma_j = +1$. The unique ground state $\sigma^0$ of $H = H_0+V$ is the all spin-up configuration, and has $W_z = +1$. The minimal energy configuration $\sigma^1$ with $W_z = -1$ is obtained from $\sigma^0$ by flipping all spins in $S_{z}$. Error membranes {\em restricted to the plane} $S_{z}$ have the same energetic cost as those in the two-dimensional Ising model $E = -2B|M|+2J|\partial M|$. At first glance, the same argument invoked for the Ising model suggests that large error membranes will proliferate in this plane and result in a configuration in the $W_z=+1$ sector, corrupting the memory.

However, at non-zero temperature, entropy will make the error membranes fluctuate in and out of the $S_{z}$ plane. Outside the plane, the magnetic field contributes positively to the energy. There is a critical temperature  where the entropy gained from fluctuating the error membrane in and out of $S_{z}$ compensates from the energy gained from restricting the error membrane to $S_{z}$. Thus, the system does not magnetize even in the presence of an external field, which is again just a restatement of Elitzur's theorem. The system with an external field is  self-dual and has been studied numerically in \cite{PhysRevB.82.085114}. We expect the self-correcting phase  to coincide with the topological phase reported in this reference.

\section{Numerical results}
\label{sec:numerics}

\subsection{Setting}

We have verified the above claims numerically. The physical process we are simulating is the following. The system is prepared at some low temperature $T_0$ with all spins up, except in the $L_z$ plane where all spins are down, so initially $W_z = -1$. With a positive magnetic field $B$, this initial configuration is not a ground state because the magnetic field favors up spins. The temperature is then slowly ramped up from $T_0$ to some holding temperature $T_{\rm hold}$. The system sits at this holding temperature for some finite amount of time $t$, after which the temperature is ramped back down to $T_0$. At that time, we measure the Wilson loop $W_z$ to check if it retained information about its initial value. Note that we do not need to run separate experiments for the two initial values of $W_z = \pm 1$ since the external field favors $W_z = +1$, i.e. the noise affecting the encoded information is asymmetric. The challenging case is when the system carries the information $W_z = -1$, and so we focus on that case.

We use standard Monte Carlo simulations with Metropolis-Hastings rule. A single Monte Carlo update implements both single spin flips and cluster updates corresponding to the gauge symmetry.  The conclusions we reach are independent of the specific choice of updates.
Each update consists of $N_s$ attempted single-spin flips, where $N_s$ is the number of spins, and $N_v$ gauge cluster updates, where $N_v$ is the number of lattice vertices.  This way, the number of Monte Carlo updates corresponds to a physical measure of time, independent of the lattice size. 
 We fix $T_0 = 0.2$. Ramping the temperature is performed with steps of $\Delta T = 0.05$. We use 5000 Monte Carlo updates at each temperature. The system is held at $T_{\rm hold}$ for 10,000 Monte Carlo sweeps, before cooling back down to $T_0$.
 Results below are averaged over 1000 such temperature sweeps.

\begin{figure}
\includegraphics[width=0.48\textwidth]{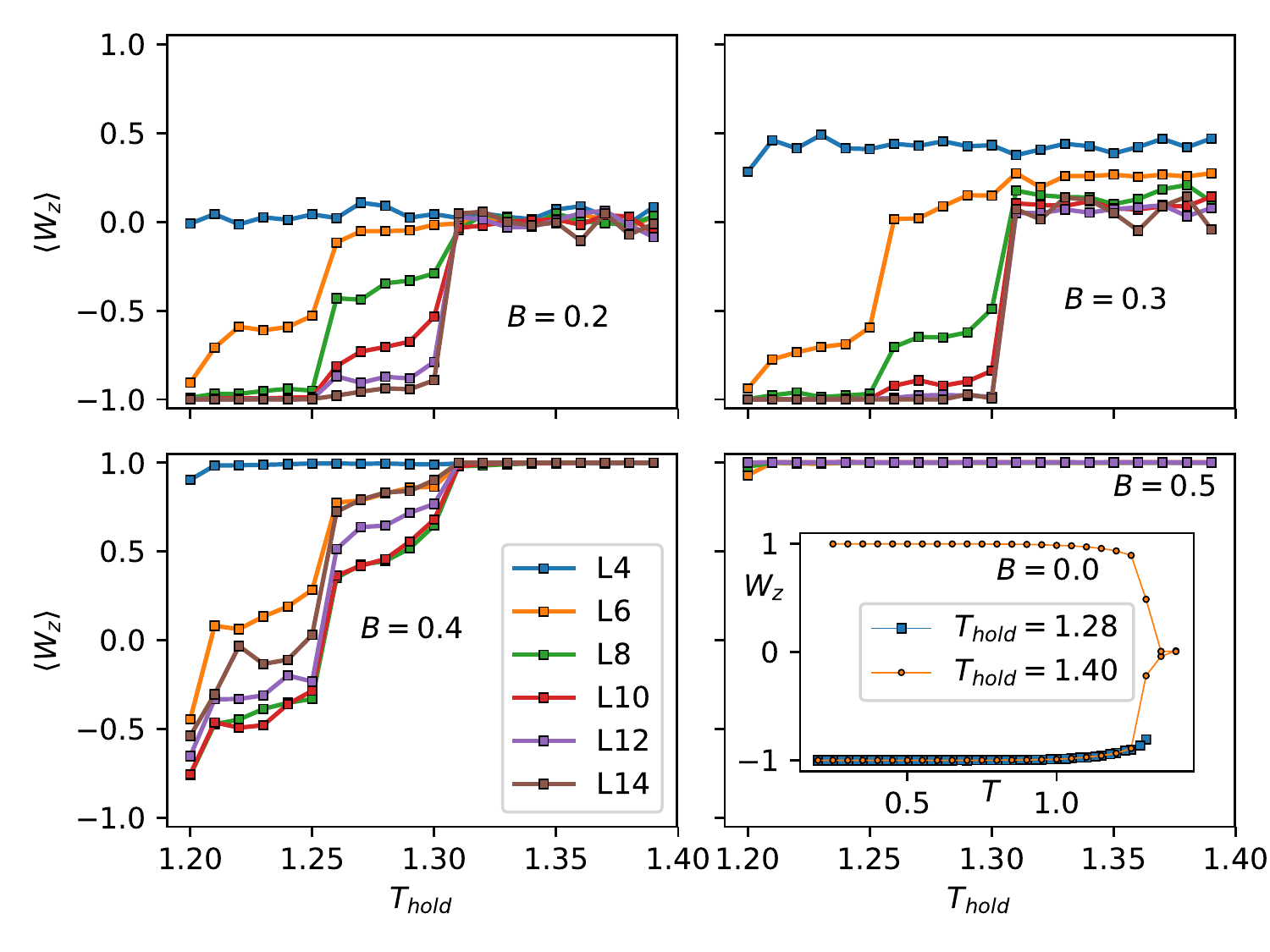}
\caption{Expectation value of the Wilson loop $\langle W_z\rangle$ as a function of the holding temperature $T_{\rm hold}$ for various system sizes and external magnetic fields $B$. For sufficiently weak external fields and sufficiently low holding temperatures, the system returns to its initial value $W_z=-1$ with a probability that increases with system size $L$, suggesting the existence of a self-correcting phase in the $(T_{\rm hold},B)$ diagram. At $B=0.4J$, we observe that the system of size $L=12$ has a higher error probability than $L=10$ for all holding temperatures, suggesting that this $B$ exceeds the critical field value.  The inset illustrates one typical temperature sweep for two different holding temperatures for a system of size $L=12$.  Each data point in the main plots correspond to an average of 1000 such temperature sweeps. 
}
\label{fig:BT}
\end{figure}

\subsection{Self-correction}

Simulation results are shown  \fig{BT} where we report the final expectation value of the Wilson loop as a function of the holding temperature for various system sizes and external fields. 
We emphasize that the main plot is not the expectation values of the Wilson loop as a function of temperature during a ramping process. Each data point is measure at $T_0$ but the measurement is preceded by a temperature ramp-up to some holding temperature and a ramp-down, and the data is plotted as a function of the holding temperature. An example of a single temperature sweep is illustrated in the inset for $B=0$. The inset shows strongly polarized values of $\langle W_z\rangle$, in apparent conflict with our claim that should vanish as $\langle W_z\rangle \approx e^{-\gamma L}$ at any finite temperature, but this is due to finite size effects since $\gamma \approx -\log(1-e^{-4J/T}) \approx 0.047$ is quite small.

\begin{figure}[t!]
\includegraphics[width=0.4\textwidth]{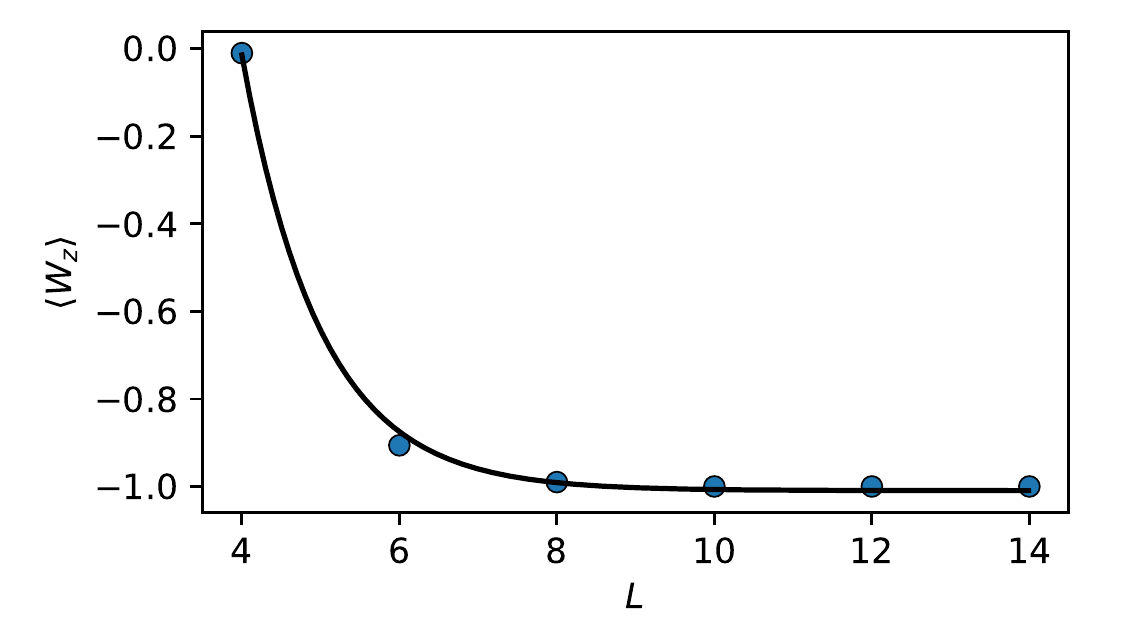}
\caption{Final Wilson loop expectation as a function of the system size for sub-critical holding temperature $T_{\rm hold} = 1.2J < T_c$ and external field $B=0.2J<B_c$. The result agrees with a fit $\langle W_z \rangle =  a\exp(-bL)-1$, shown in the solid line.}
\label{fig:exp}
\end{figure}

We say that the system with parameters $(T_{\rm hold},B,J)$ is in the self-correcting phase when the probability that the system returns to a different value of $W_z$ decreases exponentially with the system size $L$. We can verify this claim by fitting $\langle W_z\rangle$ as a function of the system size in the sub-critical region, which is shown at \fig{exp}. 

Based on these results, we predict the existence of an upper critical temperature $T_c \approx 1.31 J$ and an upper critical field $0.3 J \lesssim B_c \lesssim 0.4 J$ such that the system is self-correcting for $T_{\rm hold}<T_c$ and $B<B_c$.  The critical temperature is largely insensitive to the magnetic field until a paramagnetic phase is obtained.  These observations are consistent with the well established  critical temperature $T_c \approx 1.314J$ at $B=0$ \cite{GT96a} and more generally with the findings of \cite{PhysRevB.82.085114} which shows an upper critical field at $B_c \approx 0.225T$ which is $0.29J$ at $T=T_c$, in agreement with our observation. 

\subsection{Lower critical temperature?}

The 3D toric code is known to be a self-correcting quantum memory \cite{Kit03a,DKLP02a}. Its Hamiltonian $H_Q = H+K$ is the sum of  Wegner' classical gauge model \eq{H} and a quantum kinetic terms $K$ that commute with $H$. Because of this commutation, the partition function factors $Z_\beta(H_Q) = Z_\beta(H)\times Z_\beta(K)$, so in this case, the relations between the thermal properties of the quantum and classical model are expected. But in the presence of a magnetic field, this factorization breaks and the thermal stability of the classical model is not a direct consequence of the stability of the quantum model.

\begin{figure}[t!]
\includegraphics[width=0.4\textwidth]{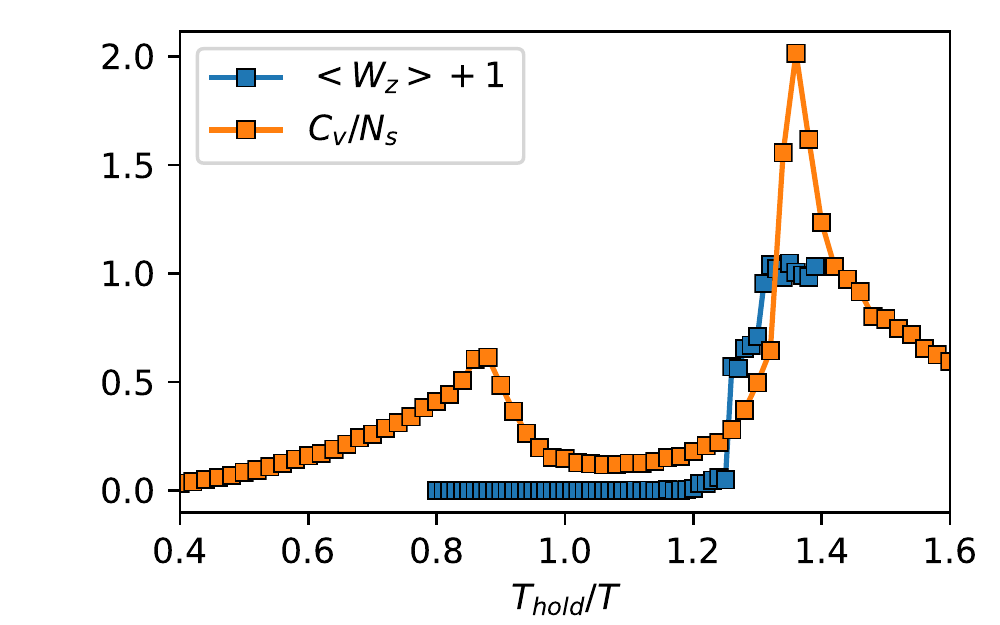}
\caption{
A comparison of the final Wilson loop expectation value for $B=0.2J$ and $L=8$ (compare to Fig.~\ref{fig:BT}) and the specific heat per spin, calculated as a typical thermodynamic estimator, for the same parameter values.
}
\label{fig:Cv}
\end{figure}

In particular, the quantum model is protected from an external magnetic field by the kinetic term, which gives it a spectral gap \cite{BHM10a,K10c,BH11b}. The classical model is only protected by entropy, an effect that we may term ``topological order by disorder". For this reason, and as discussed in \sec{GS}, we expect stability only at non-zero temperature. If we consider for instance the case of $B=0.2J$, the phase diagram proposed in \cite{PhysRevB.82.085114} predicts a lower critical temperature of $T_\ell \approx 0.89J$. However, we do not observe this lower critical temperature when simulating the system's ability to self-correct. Figure \ref{fig:Cv} shows the expectation value of the Wilson loop as a function of the holding temperatures and the specific heat as a function of temperature. While the phase transition is clearly visible on the specific head and coincides with the predictions of \cite{PhysRevB.82.085114}, there is no sign of a transition on the self-correction data. 

This discrepancy is not necessarily a problem. The phase diagram represents the equilibrium properties of a system, while self-correction is truly about the equilibration process itself. It is possible that the equilibration process remains exponentially slow below the lower critical temperature. It is also possibly that the character of the equilibration process changes at this transition, perhaps transitioning from an exponential lifetime to a polynomial lifetime which is difficult to distinguish from the system sizes we have simulated. In any case, it would be premature based on these simulations to conclude that the system remains self-correcting all the way to $T=0$, and this question deserves further scrutiny.

\section{Conclusion}

We have argued that a self-correcting memory can be realized in a lattice system with bounded degrees of freedom and local interactions if the Hamiltonian is endowed with a gauge symmetry and is embedded on a topologically non-trivial manifold. The gauge symmetry implies that the Hamiltonian spectrum is exponentially degenerate. The non-trivial topology gives rises to additional degeneracy that can be probed by a homologically non-trivial Wilson loop. Information can be reliably stored in this Wilson loop and Elitzur's theorem tells us that the stored information will not respond to an external field.  

We have tested these predictions numerically on Wegner's 3D Ising lattice gauge theory with an external magnetic field. Our results are consistent with the existence of phase in the temperature/field diagram where the information is exponentially well retained as a function of the system size, in agreement with the model's previously established phase diagram. 

The argument for the stability of the model to external fields relies on entropy, so is only expected to hold for some non-zero temperature. A lower critical temperature has been observed in previous conventional characterizations of the model, but is not seen in our numerical study. This leaves open the question of whether temperature is necessary to stabilize the encoded information. 

\medskip
\noindent{\em Acknowledgements ---} We thank Charlie Bennett for stimulating discussions. This work is supported by NSERC,
the Perimeter Institute for Theoretical Physics, and the Shared Hierarchical Academic Research Computing Network (SHARCNET). Research at Perimeter Institute is supported through Industry Canada and by the Province of Ontario through the Ministry of Research \& Innovation.

\bibliographystyle{/Users/dpoulin/archive/hsiam}

\begin{thebibliography}{10}

\bibitem{AB97a}
{\sc D.~Aharonov and M.~{Ben-Or}}, {\em Fault-tolerant quantum computation with
  constant error rate}, in Proc. 29th. Ann. ACM Symp. on Theory of Computing,
  1997, quant-ph/9611025.
\newblock Longer version quant-ph/9906129.

\bibitem{Bac05a}
{\sc D.~Bacon}, {\em Operator quantum error-correcting subsystems for
  self-correcting quantum memories}, Phys. Rev. A, 73 (2006), p.~012340.

\bibitem{BG85a}
{\sc C.~H. Bennett and G.~Grinstein}, {\em Role of irreversibility in
  stabilizing complex and nonergodic behavior in locally interacting discrete
  systems}, Phys. Rev. Lett., 55 (1985), p.~657.

\bibitem{BCG82a}
{\sc E.~R. Berlekamp, J.~H. Conway, and R.~K. Guy}, {\em Winning ways for your
  mathematicl phays}, Academic Press (New York), 1982.

\bibitem{BH11b}
{\sc S.~Bravyi and M.~Hastings}, {\em A short proof of stability of topological
  order under local perturbations}, Communications in Mathematical Physics, 307
  (2011), pp.~609--627.

\bibitem{BHM10a}
{\sc S.~Bravyi, M.~B. Hastings, and S.~Michalakis}, {\em Topological quantum
  order: stability under local perturbations}, J. Math. Phys., 51 (2010),
  p.~093512, arXiv:1001.0344.

\bibitem{B16b}
{\sc C.~G. Brell}, {\em A proposal for self-correcting stabilizer quantum
  memories in 3 dimensions (or slightly less)}, New J. Phys., 18 (2016),
  p.~013050.

\bibitem{DKLP02a}
{\sc E.~Dennis, A.~Kitaev, A.~Landahl, and J.~Preskill}, {\em Topological
  quantum memory}, J. Math. Phys., 43 (2002), p.~4452, quant-ph/0110143.

\bibitem{DO77a}
{\sc R.~L. Dobrushin and S.~I. Ortyukov}, {\em Upper bounds on the redundancy
  of self-correcting arrangements of unreliable elements}, Problems of Inform.
  Trans., 13 (1977), p.~201.

\bibitem{Eli75a}
{\sc S.~Elitzur}, {\em Impossibility of spontaneously breaking local
  symmetries}, Phys. Rev. D, 12 (1975), p.~3978.

\bibitem{Gac86a}
{\sc P.~G{\'a}cs}, {\em Reliable computation with cellular automata}, J.
  Comput. Syst. Sci., 32 (1986), pp.~15--78.

\bibitem{G88a}
{\sc P.~G{\'a}cs}, {\em A simple three-dimensional real-time reliable cellular
  array}, J. Comput. Syst. Sci., 36 (1988), p.~125.

\bibitem{G01b}
{\sc P.~G{\'a}cs}, {\em Reliable cellular automata with self-organization},
  Journal of Statistical Physics, 103 (2001), pp.~45--267.
\newblock 10.1023/A:1004823720305.

\bibitem{G61a}
{\sc J.~W. Gibbs}, {\em The Scientific Papers}, Dover Publications (New York),
  1961.

\bibitem{GT96a}
{\sc R.~Gupta and P.~Tamayo}, {\em The critical exponent for the 3d ising
  model}, US-Japan Bilateral Seminats - Maui,  (1996), p.~28.

\bibitem{HP10a}
{\sc J.~Haah and J.~Preskill}, {\em Logical operator tradeoff for local quantum
  codes}, Phys. Rev. A, 86 (2012), p.~032308, arXiv:1011.3529.

\bibitem{HCC08a}
{\sc A.~Hamma, C.~Castelnovo, and Chammon}, {\em Topological quantum memory at
  finite temperature}, 2008, 0812.4622.

\bibitem{HWRL12a}
{\sc A.~Hutter, J.~R. Wootton, B.~R{\"o}thlisberger, and D.~Loss}, {\em
  Self-correcting quantum memory with a boundary}, Physical Review A, 86
  (2012), p.~052340.

\bibitem{Kit03a}
{\sc A.~Y. Kitaev}, {\em Fault-tolerant quantum computation by anyons}, Ann.
  Phys., 303 (2003), p.~2, quant-ph/9707021.

\bibitem{K10c}
{\sc I.~Klich}, {\em On the stability of topological phases on a lattice},
  Annals of Physics, 325 (2010), pp.~2120--2131.

\bibitem{Kni05a}
{\sc E.~Knill}, {\em Quantum computing with realistically noisy devices},
  Nature, 434 (2005), p.~39, quant-ph/0410199.

\bibitem{KLZ98a}
{\sc E.~Knill, R.~Laflamme, and W.~H. Zurek}, {\em Resilient quantum
  computation}, Science, 279 (1998), p.~342.

\bibitem{Kog79a}
{\sc J.~B. Kogut}, {\em An introduction to lattice gauge theory and spin
  systems}, Rev. Mod. Phys., 51 (1979), p.~659.

\bibitem{LP13a}
{\sc O.~Landon-Cardinal and D.~Poulin}, {\em Local topological order inhibits
  thermal stability in {2D}}, Phys. Rev. Lett., 110 (2013), p.~090502,
  arXiv:1209.5750.

\bibitem{LYPP15a}
{\sc O.~Landon-Cardinal, B.~Yoshida, J.~Preskill, and D.~Poulin}, {\em
  Perturbative instability of quantum memory based on effective long-range
  interactions}, Phys. Rev. A, 91 (2015), p.~032303, arXiv:1501.04112.

\bibitem{M14a}
{\sc K.~P. Michnicki}, {\em 3d topological quantum memory with a power-law
  energy barrier}, Phys. Rev. Lett., 113 (2014), p.~130501.

\bibitem{PHWL12a}
{\sc F.~Pedrocchi, A.~Hutter, J.~Wootton, and D.~Loss}, {\em Local 3d spin
  hamiltonian as a thermally stable surface code}, 2012, arXiv:1209.5289.

\bibitem{P85a}
{\sc N.~Pippenger}, {\em On networks of noist gates}, in Proc. of the 26th IEEE
  FOCS Synposium, 1985, p.~30.

\bibitem{PB94a}
{\sc M.~Plischke and B.~Bergersen}, {\em Equilibrium statistical physics},
  World Scientific, Singapore, 1994.

\bibitem{Pre98a}
{\sc J.~Preskill}, {\em Reliable quantum computers}, Proc. R. Soc. Lond. A, 454
  (1998), p.~385, quant-ph/9705031.

\bibitem{Sho96a}
{\sc P.~W. Shor}, {\em Fault-tolerant quantum computation}, in Proceedings of
  the 37th Symposium on the Foundations of Computer Science, Los Alamitos,
  California, 1996, IEEE press, pp.~56--65, quant-ph/9605011.

\bibitem{S96b}
{\sc D.~Spielman}, {\em Highly fault-tolerant parallel computation}, in Proc.
  of the 37th IEEE FOCS Synposium, 1996, p.~154.

\bibitem{Ste96c}
{\sc A.~M. Steane}, {\em Simple quantum error correcting codes}, Phys. Rev. A,
  54 (1996), p.~4741.

\bibitem{T80a}
{\sc A.~Toom}, {\em Stable and attractive trajectories in multicomponent
  systems}, in Multicomponent Systems, R.~L. Dobrushin, ed., vol.~6 of Advanced
  in Probability, Dekker (New York), 1980, p.~549.

\bibitem{PhysRevB.82.085114}
{\sc I.~S. Tupitsyn, A.~Kitaev, N.~V. Prokof'ev, and P.~C.~E. Stamp}, {\em
  Topological multicritical point in the phase diagram of the toric code model
  and three-dimensional lattice gauge higgs model}, Phys. Rev. B, 82 (2010),
  p.~085114.

\bibitem{vN56a}
{\sc J.~von Neumann}, {\em Probabilistic logics and the synthesis of reliable
  organisms from unreliable components}, in Automata Studies, Princeton, NJ,
  1956, Princeton University Press, pp.~329--378.

\bibitem{W71a}
{\sc F.~Wegner}, {\em Duality in generalized {Ising} models and phase
  transitions without local order parameters}, J. Math. Phys., 12 (1971),
  p.~2259.

\end{thebibliography}

\end{document}